\documentclass[aps,prl,twocolumn,superscriptaddress,groupedaddress]{revtex4}

\usepackage{mathrsfs}
\usepackage{graphicx}
\usepackage{bm}        
\usepackage{amssymb}   
\usepackage{amsmath}
\usepackage{color}
\usepackage{subfigure} 

\def\bea{\begin{eqnarray}}
\def\eea{\end{eqnarray}}

\def\be{\begin{equation}}
\def\ee{\end{equation}}

\begin{document}


\title{Insight into the Unwrapping of the Dinucleosome}

\author{Fatemeh Khodabandeh}%
\affiliation{Department of Physics, Institute for Advanced Studies in
Basic Sciences (IASBS), Zanjan 45137-66731, Iran}

\author{Hashem Fatemi}%
\affiliation{Department of Physics, Institute for Advanced Studies in
Basic Sciences (IASBS), Zanjan 45137-66731, Iran}

\author{Farshid Mohammad-Rafiee}%
\affiliation{Department of Physics, Institute for Advanced Studies in
Basic Sciences (IASBS), Zanjan 45137-66731, Iran}

\date{\today}


\begin{abstract}
Dynamics of nucleosomes, the building blocks of the chromatin, has crucial effects on expression, replication and repair of genomes in eukaryotes. Beside constant movements of nucleosomes by thermal fluctuations, ATP-dependent chromatin remodelling complexes cause their active displacements. Here we propose a theoretical analysis of dinucleosome wrapping and unwrapping dynamics in the presence of an external force. We explore the energy landscape and configurations of dinucleosome in different unwrapped states. Moreover, using a dynamical Monte-Carlo simulation algorithm, we demonstrate the dynamical features of the system such as the unwrapping force for partial and full wrapping processes. Furthermore, we show that in the short length of linker DNA ($\sim 10 - 90$ bp), the asymmetric unwrapping occurs. These findings could shed some light on chromatin dynamics and gene accessibility.  

\end{abstract}

\maketitle

\section*{INTRUDUCTION}

Chromatin has a crucial role in the essential eukaryotic biological processes such as replication, transcription, recombination, and other gene regulations. The basic subunit of chromatin is the nucleosome core particles (NCP), which consists of 147 base pair (bp) DNA wrapped around the core of histone proteins \cite{Cell}. In chromatin, these NCPs are separated from each other by variable lengths of linker DNA ($ \sim 10 - 100$ bp) \cite{Widom-1997,Khorasan-2004}. Since a typical gene is composed of tens to hundreds of nucleosomes, the accessibility to gene contents is vital for the life cycle of cells. 

{\it In vivo} and {\it in vitro} experiments showed that the gene contents of chromatin are accessible due to the structural fluctuations of the nucleosomal DNA \cite{Widom-2000,Flaus-2003,Widom-2005}. This dynamical behavior becomes more important and interesting when active translocation motors such as RNA polymerases or remodelers exert effective forces and torques on the nucleosomes \cite{Bustamante-2009,Logie-2009}. In these situations, one may ask about the effect of force on the nucleosome dynamics. 

Using optical and magnetic tweezers, one can apply a tunable force on a single nucleosome or a nucleosome array. In the experiments on the nucleosome arrays, an external force has been applied, and the unwrapping transition at high force \cite{Bennink-2001,Wang-2002} and low force limit \cite{Bustamante-2000} has been examined. In these studies, the details of mononucleosome unwrapping transition were not clear. 
Recently, some experiments have been done for studying the unfolding dynamics of the 30-nm chromatin fiber at the low force regime \cite{vanNoort-Nat2009}. Although the elastic properties of the 30-nm chromatin fiber depend on its architecture \cite{vanNoort-NAR2015}, it has been shown that when the stretching force is smaller than $\sim 4$ pN, the fiber behaves similarly to a Hookian spring, whereas its stiffness and the force-extension relation suggest that the nucleosomes in the fiber are organized in a one-start solenoidal architecture \cite{vanNoort-Nat2009}. 

Understanding the dynamics of a single nucleosome under external tension was the subject of experimental investigations in the past decade \cite{Mihardja-2006,vanNoort-2009-2,Mack-2012,Wang-2013}. These experiments revealed that the nucleosomal DNA unwinds from the histone octamer in two main stages, where at the first stage, the first turn of the nucleosomal DNA is unwound and in the last stage the second half of the wrapped portion is opened. The interesting point is that the first turn is opened mostly reversible, and the force needed to open it is about 3 pN. The second turn is unwound in larger forces widely distributed around $\sim 8-9$ pN, and its opening is irreversible on the experimental timescales. Beside advanced experimental setups, theoretical descriptions of such a dynamical system have been developed in the past years, taking into account different DNA-histone binding energies on binding sites \cite{Wang-2002,Wang-2009}, force-induced unwrapping considering the effect of bent DNA close to the entry-exit points of the nucleosome and the electrostatic repulsion between two turns of the nucleosomal DNA \cite{Kulic-2004}, dynamical models with considering the fluctuations of the DNA \cite{Spakowitz-2011}, and with considering fourteen binding sites in the nucleosome and more details of nucleosomal DNA \cite{Laleh-2012}.  Recently, by combining fluorescence with optical tweezers, it has been observed that when the stiffness of the nucleosomal DNA changes along its length, the nucleosome can unwrap asymmetrically in the presence of external force \cite{Ngo-2015}. In addition, it has been noticed that the unwrapping of the nucleosomal DNA from the stiffer side is most likely.   

Here, we study the dynamics of the unwrapping and rewrapping of a dinucleosome under an external force. The energy of the system and the conformations of the dinucleosome are determined using a simple elastic model based on Ref. \cite{Hashem-Fatemeh1}, whereas the energy landscape is used in the dynamical Monte-Carlo simulation for studying the dynamics of the problem. We see that for the forces larger than $\sim 3$ pN, the most likely situation is that the first turn of one of the nucleosomes will be unwrapped and after that, the first turn of the second nucleosome is unwrapped. Furthermore, we see the length of the linker DNA plays an important role in the problem: for long linker DNA lengths the details of the unwrapping-rewrapping of the nucleosomes are not important, but for short enough linker DNA lengths, the dinucleosome system prefers to unwrap asymmetrically. 


\section*{MATERIALS AND METHODS}

\begin{figure}
 \centering
 \includegraphics[width=0.99\linewidth]{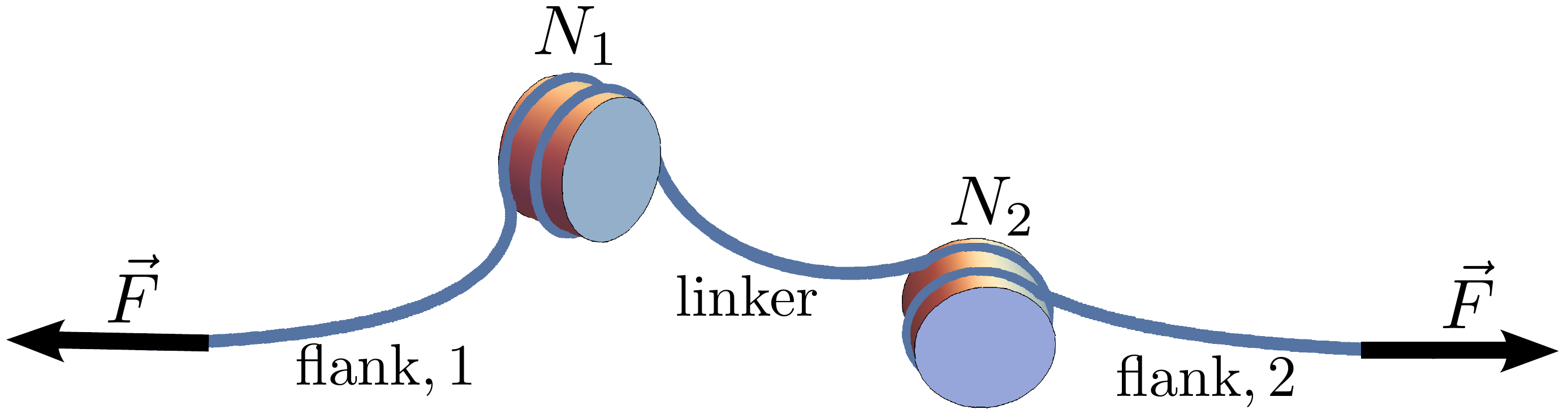}
 \caption{The schematic picture of a dinucleosome with two flanking DNA. The nucleosomes are denoted by $N_1$ and $N_2$. The plot is corresponding to the situation that the first and the second nucleosomes are in states $n_1=1$ and $n_2 = 6$, respectively. The applied force is in the direction of $z$.}
 \label{fig:schematic}
\end{figure}

We consider two nucleosomes on a long DNA under an external stretching force and study the wrapping and unwrapping dynamics of the nucleosomes. We assume that the two flanking DNAs are long enough and there is a linker DNA between the two histone octamers with the length $L_{\rm linker}$. The schematic picture of the problem has been shown in Fig. \ref{fig:schematic}. In order to study the wrapping-unwrapping dynamics of the nucleosomes, we need to have the energy landscape of the problem. We calculate the energy of the dinucleosome in the force and torque balance condition using the model that has been described in Ref. \cite{Hashem-Fatemeh1}. In this model, the DNA is considered as an elastic slender rod with the bending rigidity of $\kappa$. We assume that any changes in the twist of DNA can be washed out in the boundaries of flanking DNAs \cite{Hashem-Fatemeh1,Nam-Arya-2014}. X-ray crystallography experiments reveal that there are 14 binding sites in nucleosomes, where the minor grooves of the DNA face inwards to the histone proteins, and the DNA length between two adjacent binding sites is about 10 bp \cite{Luger-1997,Richmond-2003}. We introduce $n_{i}$ as the number of opening binding sites for the nucleosome $i$. In general, the total energy of the dinucleosome depends on the external force, $F$, the length of the linker DNA, $L_{\rm linker}$, the lengths of two flanking DNAs, $L_{\rm flank,1}$ and $L_{\rm flank,2}$, and the number of opening binding sites of the nucleosomes, $n_1$ and $n_2$, and can be written as
\bea
E_{tot} &=& E_{\rm Nuc,1}(n_1) + E_{\rm Nuc,2}(n_2)  \nonumber \\ 
&+& E_{\rm linker}(F,L_{\rm linker}, n_1, n_2) \nonumber \\
&+& E_{\rm flank,1}(F, L_{\rm flank,1},n_1) +  E_{\rm flank,2}(F, L_{\rm flank,2},n_2). \nonumber \\  
\label{eq:E-general}
\eea

As can be seen in Fig. \ref{fig:schematic}, there are two segments of the DNA wrapped around the histone octamers, which will be called ``nucleosomal DNA'', and three segments of the DNA which are free. The energy of the nucleosomal DNA consists of three parts: (1) the elastic energy of the deformed DNA, (2) the DNA-histone adsorption interactions, and (3) DNA-DNA electrostatic repulsive interactions. The two first contributions can be considered as effective adsorption energy per binding sites and are denoted by $\varepsilon_{ads}$. The DNA-DNA repulsion energy per unit length is indicated by $\varepsilon_{es}$. We note that this term is considered when the wrapping nucleosomal DNA has more than one turn. Therefore, the total energy of the nucleosomal DNA for the nucleosome $i$ can be written as \cite{Laleh-2012}
\bea
E_{\rm Nuc,i}(n_i) = a n_i \left[ \varepsilon_{ads} + \varepsilon_{el} \, \Theta(7-n_i) \right], \label{eq:E-Nuc}
\eea
where $a = 10$ bp is the mean length of the DNA between two adjacent binding sites and $\Theta(x)$ shows the step function that is zero for $x<0$, and 1 for $x \ge 0$. It should be noted that in the physiological conditions, the DNA prefers to wrap around the octamer and therefore $\varepsilon_{ads}$ has a negative value. Furthermore,  when $n_i$ is less than 7, there is an effective repulsive energy due to DNA-DNA electrostatic interactions and $\varepsilon_{el}$ should have a positive value. 

There are two contributions to the energy of the linker and the flanking DNAs: the elastic bending energy and the work of the external force. Considering these two contributions, the energy of each linker and flanking DNA is written as
\bea
E_{\rm DNA} = \int_0^L \frac{\kappa}{2} \left( \frac{d \hat{t}}{ds} \right)^2 ds - \int_0^L \vec{F} \cdot \hat{t} \, ds, \label{eq:elastic-energy}
\eea 
where $L$ denotes the length of the considered naked DNA, which can be the linker or the flanking part, and 
$\hat{t}$ is the local tangent unit vector of the DNA at the arclength of $s$. As we have discussed in Ref. \cite{Hashem-Fatemeh1}, there is a constraint on the local tangent vector of the DNA at the positions where the DNA exits the histone octamer. Therefore, when the nucleosomal DNA is partially unwrapped, the local tangent vectors of the free DNAs at the exit points of the histone octamers are changed. Considering the force and torque balance conditions of the initial equilibrium state, the mentioned changes at the boundaries of the free DNAs cannot hold the equilibrium conditions and so the orientations of the nucleosomes change in order to go to the equilibrium conformation. 

Therefore, when the nucleosomal DNAs are partially unwrapped, the system goes from $(n_1=0,n_2=0)$ to a new state of $(n_1,n_2)$. Due to this transition, the total energy of the system is changed. The orientation of the octamers and the conformation of the DNAs can be determined using the method described in Ref. \cite{Hashem-Fatemeh1}. The total energy of the system at the new state of $(n_1,n_2)$ is determined by Eqs. (\ref{eq:E-general})--(\ref{eq:elastic-energy}) under the force and torque balance conditions. It is possible to have more than one solution that fullfill the force and torque balance conditions for a given state of $(n_1,n_2)$. In this paper, we choose a solution corresponding to the global minimum energy for the system.

Knowing the energy landscape of the problem in terms of $(n_1,n_2)$, enables us to estimate the opening and closing rates of the binding sites of the nucleosomes. We assume that the dominant mechanism for the opening and closing of the binding sites of the nucleosomes is one-by-one. It means that at each time step we can have one of the following events: $(n_1,n_2) \rightarrow (n_1 \pm 1, n_2)$ or $(n_1,n_2) \rightarrow (n_1 , n_2 \pm 1)$. For example, when one of the binding sites of the nucleosome 1 is opened or closed, we have $(n_1,n_2) \rightarrow (n_1 + 1, n_2)$ or $(n_1,n_2) \rightarrow (n_1 - 1, n_2)$, respectively. After defining $k_{0uw}^i$ and $k_{0rw}^i$ as the unwrapping and rewrapping rates of the binding sites, respectively, at zero force for nucleosome $i$, one can write the transition rates as \cite{Laleh-2012}
\begin{widetext}
\vspace*{-0.5cm}
\begin{subequations}
\bea
k^{(1)}_{uw} (F,n_1,n_2) &=& k_{0uw}^{(1)} \, e^{-\lambda_F \beta \left[ \Delta E^*(n_1+1,n_2,F) - \Delta E^*(n_1,n_2,F)\right]} \label{eq:kuw1} \\
k^{(1)}_{rw} (F,n_1,n_2) &=& k_{0rw}^{(1)} \, e^{(1-\lambda_F) \beta \left[ \Delta E^*(n_1+1,n_2,F) - \Delta E^*(n_1,n_2,F)\right]} \label{eq:krw1} \\
k^{(2)}_{uw} (F,n_1,n_2) &=& k_{0uw}^{(2)} \, e^{-\lambda_F \beta \left[ \Delta E^*(n_1,n_2+1,F) - \Delta E^*(n_1,n_2,F)\right]} \label{eq:kuw2} \\
k^{(2)}_{rw} (F,n_1,n_2) &=& k_{0rw}^{(2)} \, e^{(1-\lambda_F) \beta \left[ \Delta E^*(n_1,n_2+1,F) - \Delta E^*(n_1,n_2,F)\right]} \label{eq:krw2},
\eea
\end{subequations}
\end{widetext}
where $\beta \equiv 1/(k_BT)$, $\lambda_F$ denotes a load distribution factor that will be considered as a fitting parameter \cite{Kolomeisky-Fisher-2000}, and $\Delta E^*$ is defined as
\bea
\begin{split}
& \Delta E^*(n_1,n_2,F)  \equiv \\
& E_{min}(n_1,n_2,F) - E_{min}(n_1,n_2,F=0).
\label{eq:E*}
\end{split}
\eea
In Eq. (\ref{eq:E*}), $E_{min}(n_1,n_2,F)$ denotes the total energy of the system (Eq. (\ref{eq:E-general})), when the dinucleosome is at equilibrium under the external force $F$, and the number of binding sites of the nucleosome $i$ is $n_i$. It is worth mentioning that the unwrapping and rewrapping rates at zero force, $k_{0uw}^i$ and $k_{0rw}^i$, can be estimated using the detailed balance equation as \cite{Hanggi-RMP-1990}

\bea
\frac{k_{0uw}^{(1)}}{k_{0rw}^{(1)}} &=& e^{-\beta \left[ E_{min}(n_1+1,n_2,F=0) - E_{min} (n_1,n_2,F=0) \right] }, \label{eq:k0-1} \\
\frac{k_{0uw}^{(2)}}{k_{0rw}^{(2)}} &=& e^{-\beta \left[ E_{min}(n_1,n_2+1,F=0) - E_{min} (n_1,n_2,F=0) \right] }. \label{eq:k0-2}
\eea

Using these transition rates, we simulate the dynamics of the system employing the Gillespie algorithm \cite{Gillespie-1976}. We use the parameters suggested in Ref. \cite{Laleh-2012}, namely $\gamma_F =0.6$, $k_{0rw}^{(i)} = 10^{4} \; {\rm s}^{-1}$, $\epsilon_{es} = 0.2  \;  {\rm k_BT/nm}$, and $\epsilon_{ads} = 0.78  \;  {\rm k_BT/nm}$. Using this set of parameters and the model described in Ref. \cite{Laleh-2012}, the simulation results for the dynamical behavior of a mono-nucleosome under external tension are in very good agreement with the experimental data. 


\section*{RESULTS}

\begin{figure}
 \centering
 \includegraphics[width=1.0\linewidth]{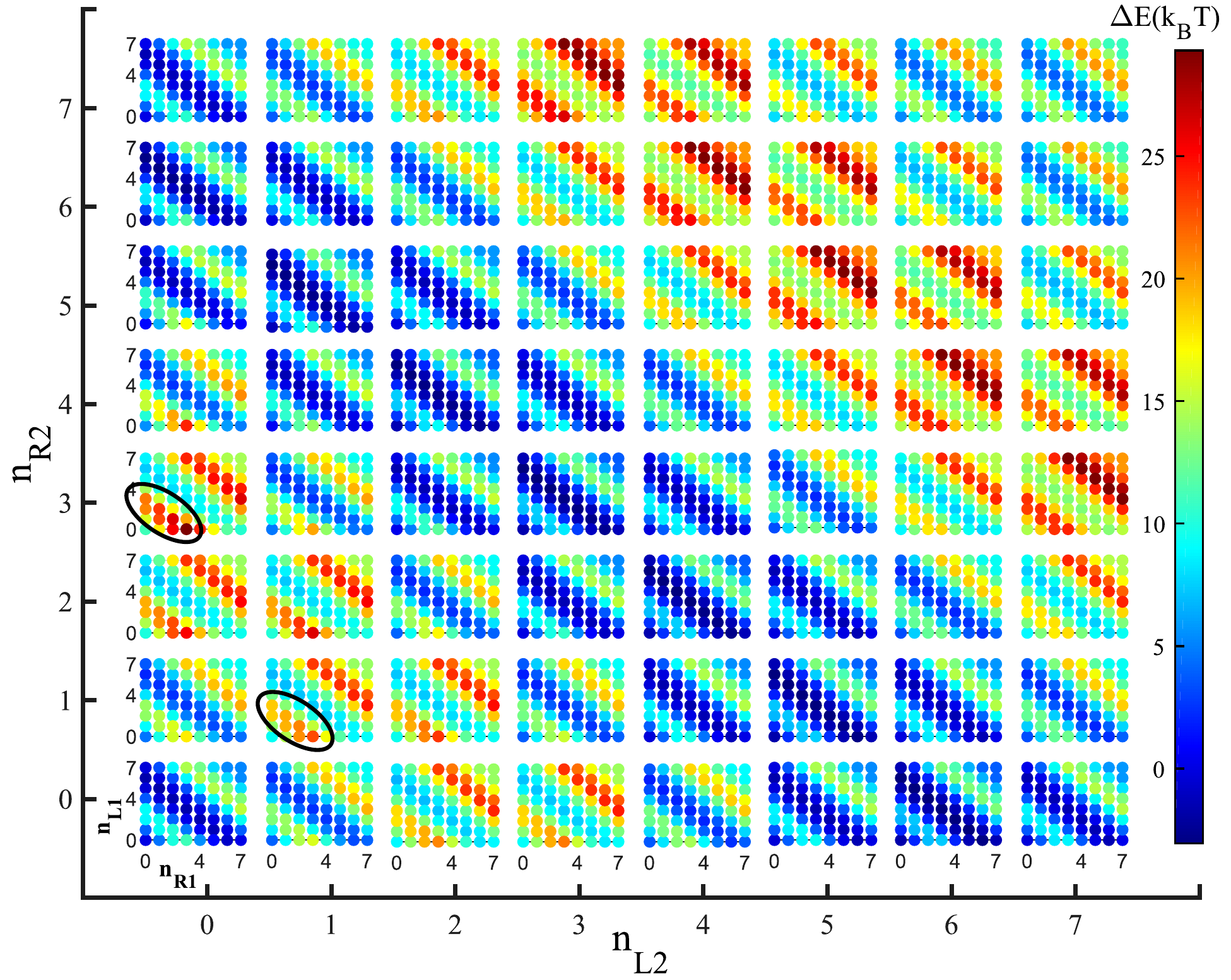} 
\caption{ (color online) The representation of $\Delta E \equiv E_{min}(n_1,n_2) - E_{min}(n_=0,n_2=0)$ vs. the number of opened binding sites of the nucleosomes for $F=3$ pN and $L_{linker}=20$ bp. $n_{Li}$ and $n_{Ri}$ represent the number of opened binding sites from the left and right of the nucleosome $i$. The color bar in the right panel shows different colors corresponding to the energy in the unit of $k_BT$.  Two examples of asymmetric regions in the unwrapping energy landscapes are shown by two ellipses.}
\label{fig:unwrapping-landscape-F3-L20} 	    
\end{figure}

The energy landscape of a mono-nucleosome is symmetric in the phase space of the number of opened binding sites from the left and right of the nucleosome, which means the symmetric and asymmetric unwrapping of the nucleosome have the same energy. One may then ask what happens in the dinucleosome case, where the length of the linker DNA may change the energy landscape of the system. As it is shown in Fig. \ref{fig:unwrapping-landscape-F3-L20}, for short lengths of the linker DNA like $L_{linker} = 20$ bp, the energy landscape of the dinucleosome unwrapping is asymmetric. The inset plots of the figure, represent the free energy landscape of the first nucleosome unwrapping, while the second nucleosome remains intact in a fixed number of opened binding sites. Two examples of asymmetric regions in the landscape of the unwrapping energy are shown by two ellipses in the figure. For example, when $\left(n_{L2}, n_{R2}, n_{R1}, n_{L1} \right) = \left( 0,3,3,0\right)$, the color of the point is dark red and the energy is  $\sim 30 \, {\rm k_BT}$, whereas $\left(n_{L2}, n_{R2}, n_{R1}, n_{L1} \right) = \left( 0,3,0,3\right)$, the color of the point is orange and the energy is  $\sim 20 \, {\rm k_BT}$. The two mentioned points are located in the left ellipse in the Fig. \ref{fig:unwrapping-landscape-F3-L20}. 

\begin{figure}
 \centering
 \includegraphics[width=0.95\linewidth]{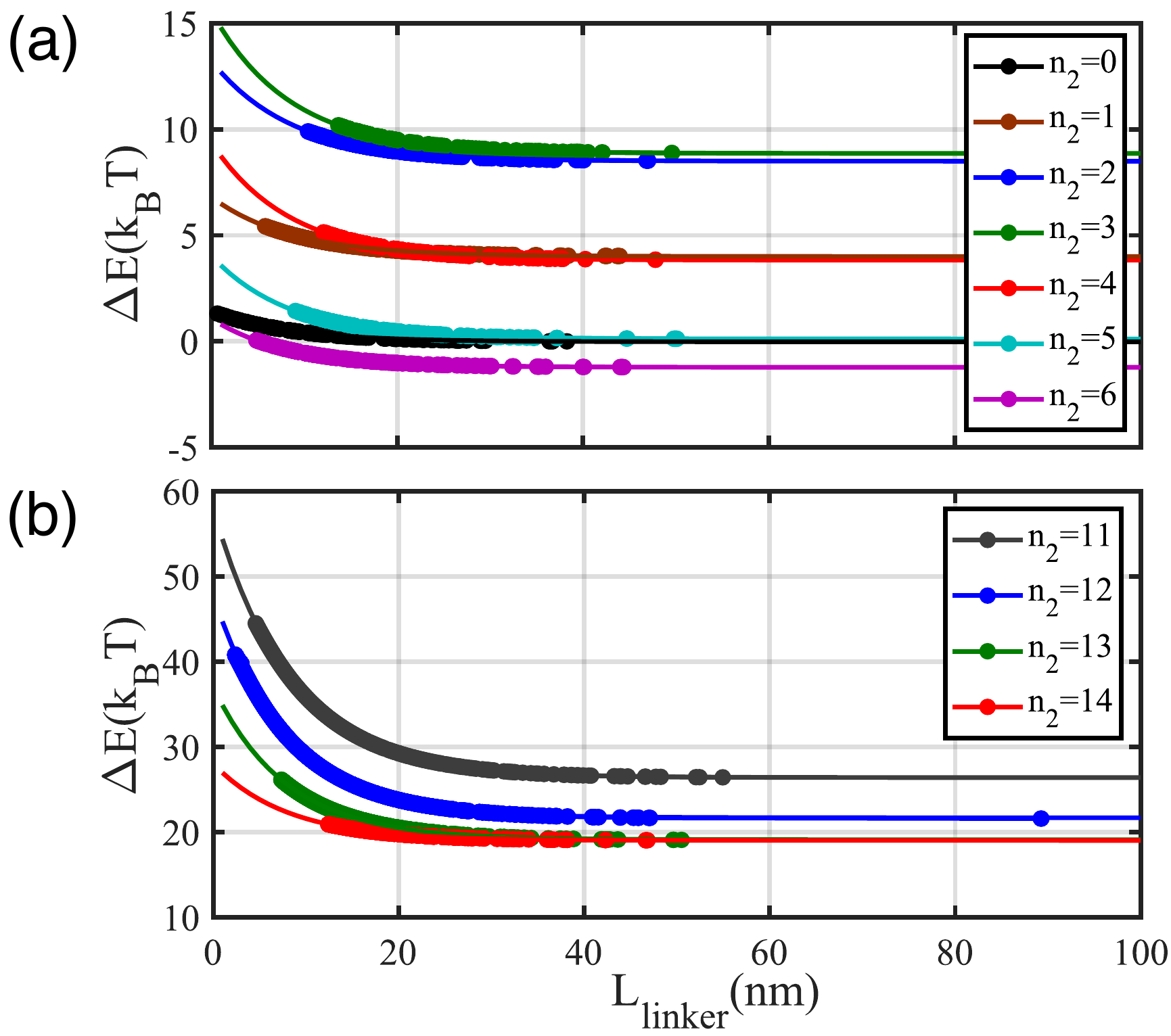} 
\caption{ (color online)  $\Delta E \equiv E_{min}(n_1,n_2,L_{linker}) - E_{min}(n_1=0,n_2=0,L_{linker}=100 \, {\rm nm})$ in terms of the length of the linker DNA for $F=3$ pN and different values of $n_2$. Plot (a) corresponds to $n_1 = 0$ and plot (b) corresponds to $n_1 = 11$, respectively. }
\label{fig:E-linker-n1n2} 	    
\end{figure}

As the number of opened binding sites of the nucleosomes and accordingly the length of the linker DNA increases, the asymmetric behavior is washed out. Unlike the fully wrapped dinucleosome \cite{Hashem-Fatemeh1}, this result is due to the dependency of the dinucleosome energy on the length of the linker DNA. To show this dependency more clearly, in Fig. \ref{fig:E-linker-n1n2}, the behavior of $\Delta E$ in terms of the length of the linker DNA is shown for different unwrapping cases. Plot (a) in Fig. \ref{fig:E-linker-n1n2} corresponds to the unwrapping of the first turn of the second nucleosome, when the first one is in the fully wrapped state. As can be seen, when the length of the linker DNA becomes longer, the difference in energy becomes smaller. We can see the same behavior, when the second turn of the second nucleosome starts to unwrap, see Fig.  \ref{fig:E-linker-n1n2}(b). When the length of the linker DNA is short, due to the structural constraint, the linker DNA should be bent dramatically; and that is why there is a significant change in the amount of the energy, when the length of the likner DNA is short.

\begin{figure}
\centering
\includegraphics[width=1\linewidth]{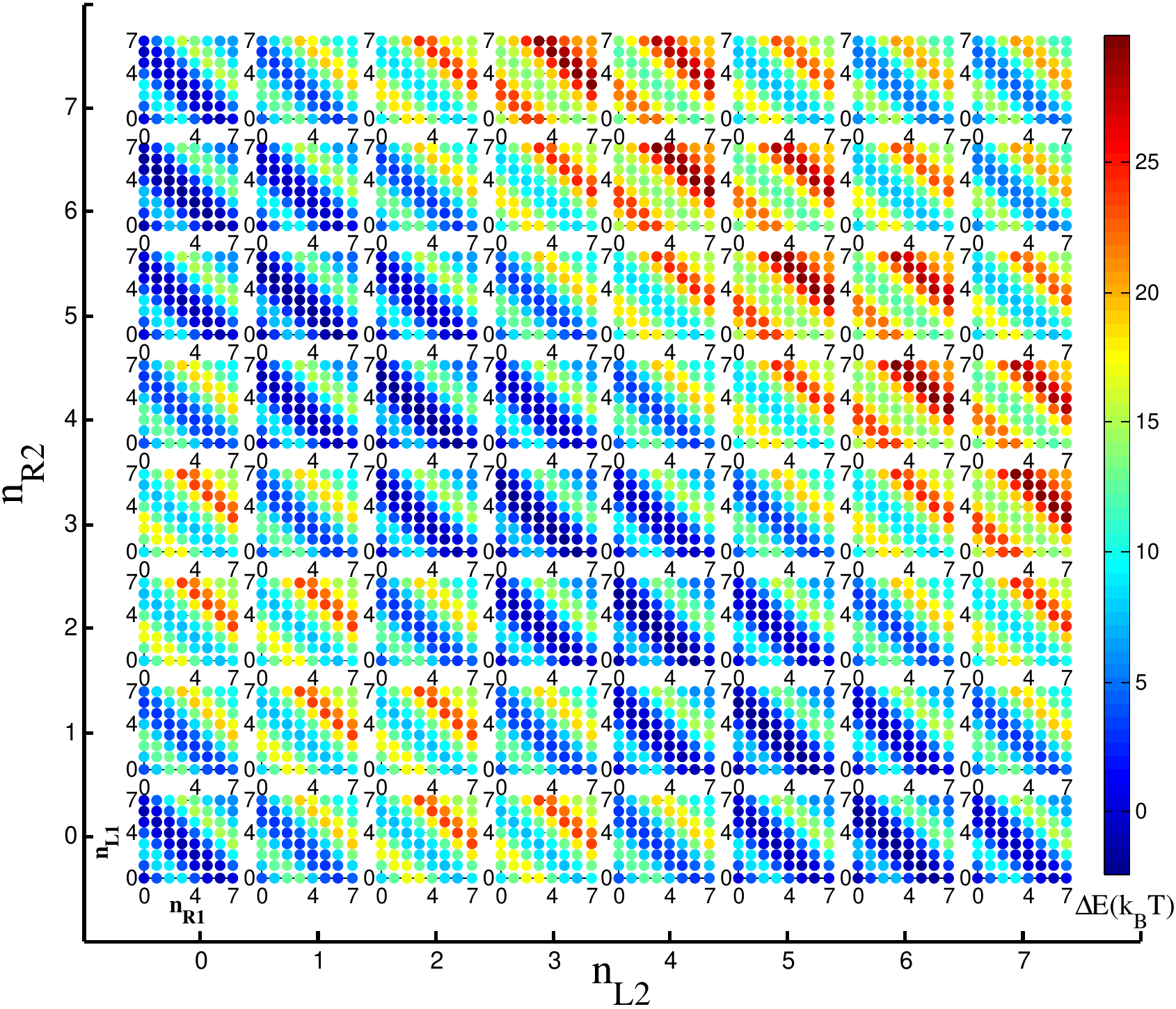}
\caption{(color online) The representation of $\Delta E \equiv E_{min}(n_1,n_2) - E_{min}(n_=0,n_2=0)$ vs. the number of opened binding sites of the nucleosomes for $F=3$ pN and $L_{linker}=200$ bp. $n_{Li}$ and $n_{Ri}$ represent the number of opened binding sites from the left and right of the nucleosome $i$. The color bar in the right panel shows different colors corresponding to the energy in the unit of $k_BT$.}
\label{fig:En_F=3_LM=200}
\end{figure}

Thus one expects that the four-dimensional energy landscape should be symmetric when the length of the linker DNA is long enough, as can be seen in Fig. 
\ref{fig:En_F=3_LM=200}. The plot corresponds to $F = 3 $ pN and $L_{linker} = 200$ bp. In this situation, the details of the nucleosomes unwrapping are not important, and the energy can be obtained as a function of $n_1$ and $n_2$, alongside $L_{Linker}$ and $F$ dependencies. 

\begin{figure}[ht]
 \centering
 \includegraphics[width=1\linewidth]{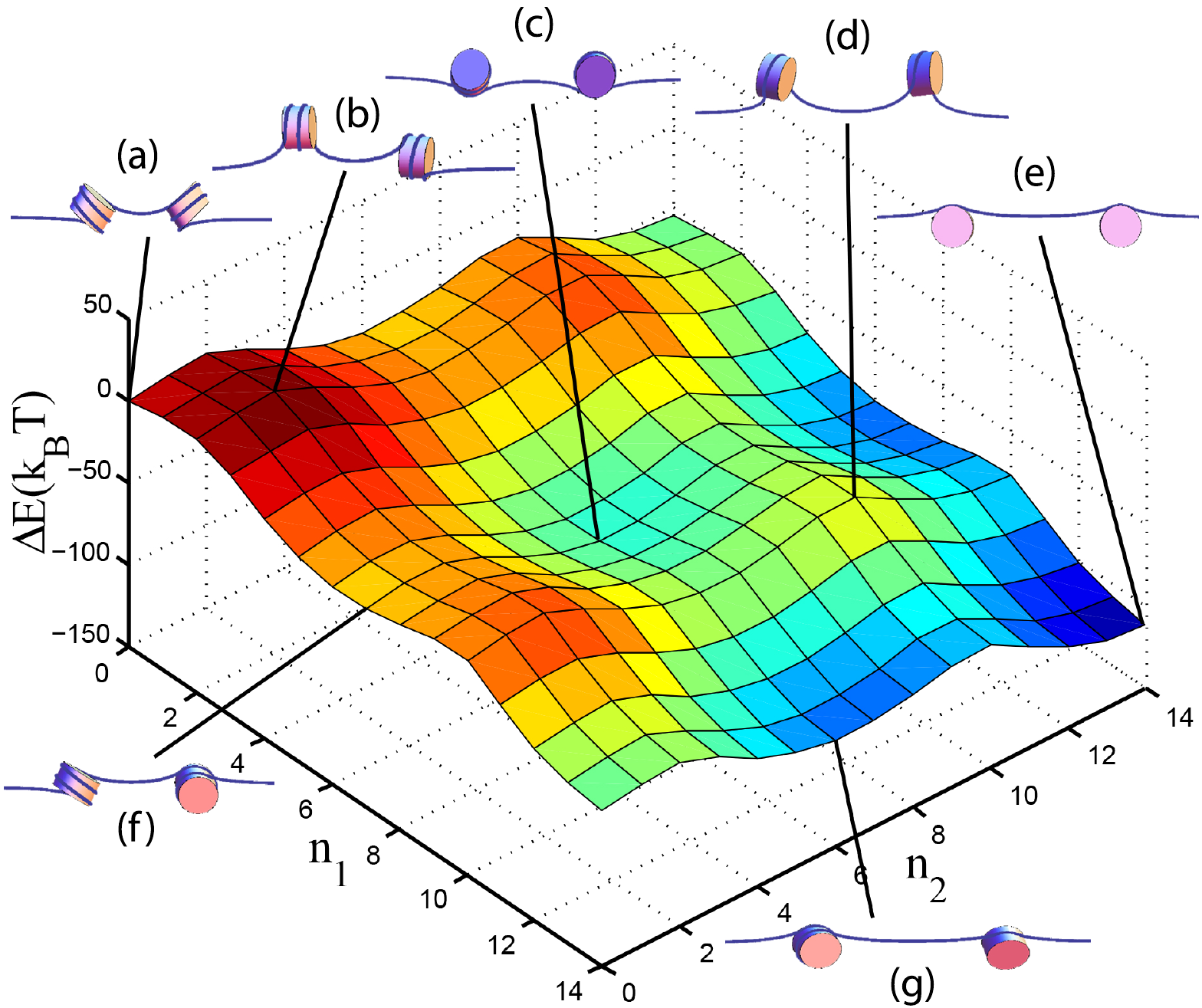}
 \caption{$\Delta E \equiv E_{min}(n_1,n_2) - E_{min}(n_=0,n_2=0)$ vs. the number of opened binding sites of the nucleosomes ($n_1,n_2$) for $F=8$ pN and $L_{linker}=200$ bp. The conformations of the nucleosomes are shown for different partially unwrapped states of nucleosomes.}
 \label{fig:E_n1n2_F8_S}
\end{figure}

As we mentioned before, the local conformation of the nucleosomes and their relative orientation respect to each other, depends on the linker DNA length, the external force, and the number of opened binding sites. Fig.  \ref{fig:E_n1n2_F8_S} shows the energy landscape of the dinucleosome in terms of the opened binding sites. Furthermore the local conformation of the two nucleosomes in different situations have been shown in the figure. The plot corresponds to $F = 8 $ pN and $L_{linker} = 200$ bp. As we discussed above, we expect the energy landscape to be symmetric in terms of $n_1$ and $n_2$ for long length of the linker DNA, which can be seen in Fig. \ref{fig:E_n1n2_F8_S}. We also expect that for long enough linker DNA lengths, the dinucleosome system behaves like two separate mono-nucleosomes. This result can be tested by comparing the energy landscape of a dinucleosome with the sum of the two mono-nucleosomes energies. We have tested this argument for different lengths of $L_{linker}$ and see that the intuition is fulfilled for long $L_{linker}$.

\begin{figure}
 \centering
 \includegraphics[width=0.95\linewidth]{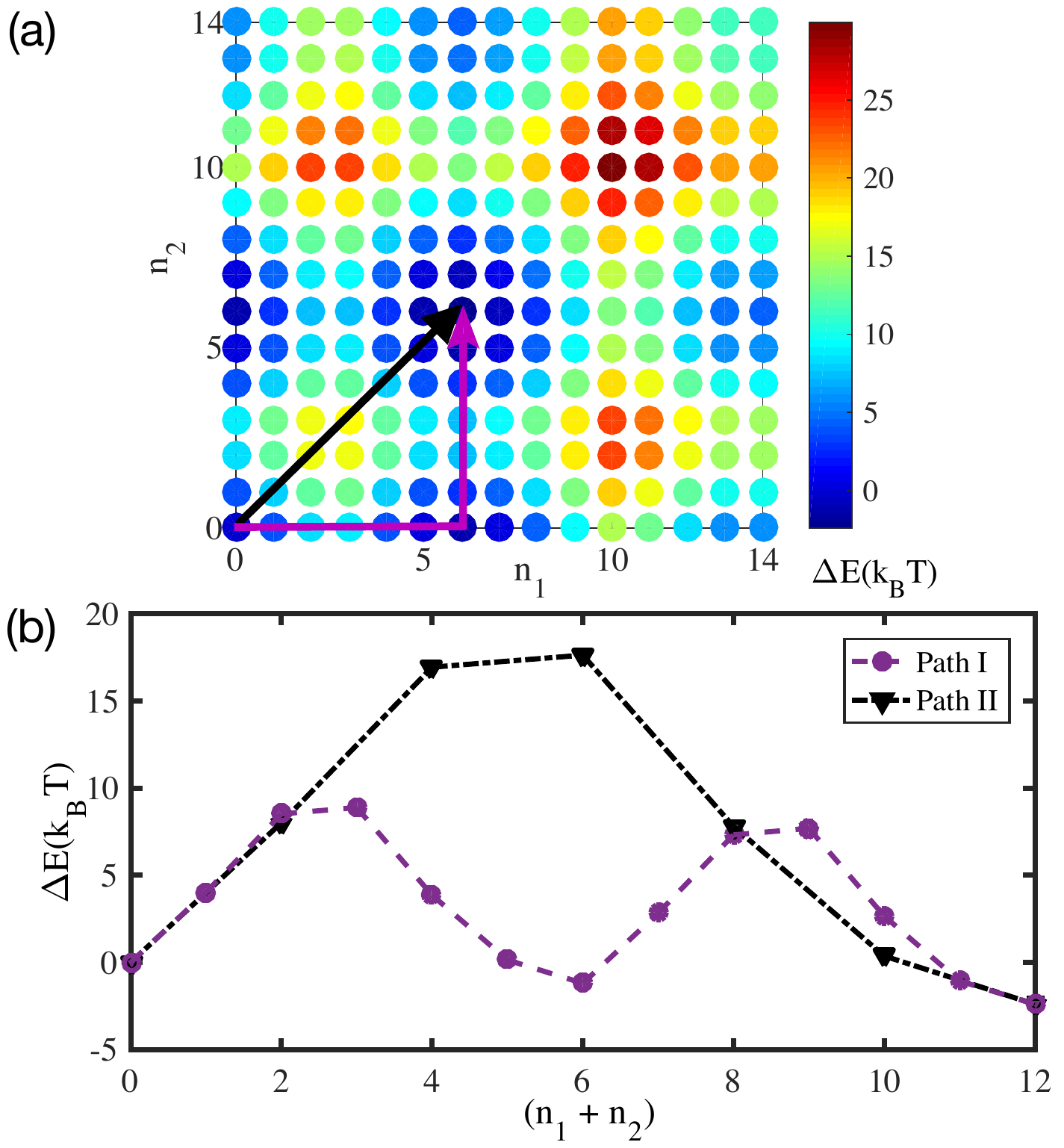}
 \caption{(a) The free energy landscape of the dinucleosome, $\Delta E \equiv E_{min}(n_1,n_2) - E_{min}(n=0,n_2=0)$, versus the number of opened binding sites of the nucleosomes. The color bar in the right panel shows different colors corresponding to the energy in the unit of $k_BT$. Two possible paths for unwrapping are shown by the purple (path I) and black (path II) arrows. (b) $\Delta E$ versus the total number of opened binding sites, $n_1+n_2$, for the two paths shown in (a). These plots correspond to $F=3$ pN and $L_{linker} = 200$ bp.  }
\label{fig:En_landscape_F3} 	    
\end{figure}

\begin{figure}
 \centering
 \includegraphics[width=0.96\linewidth]{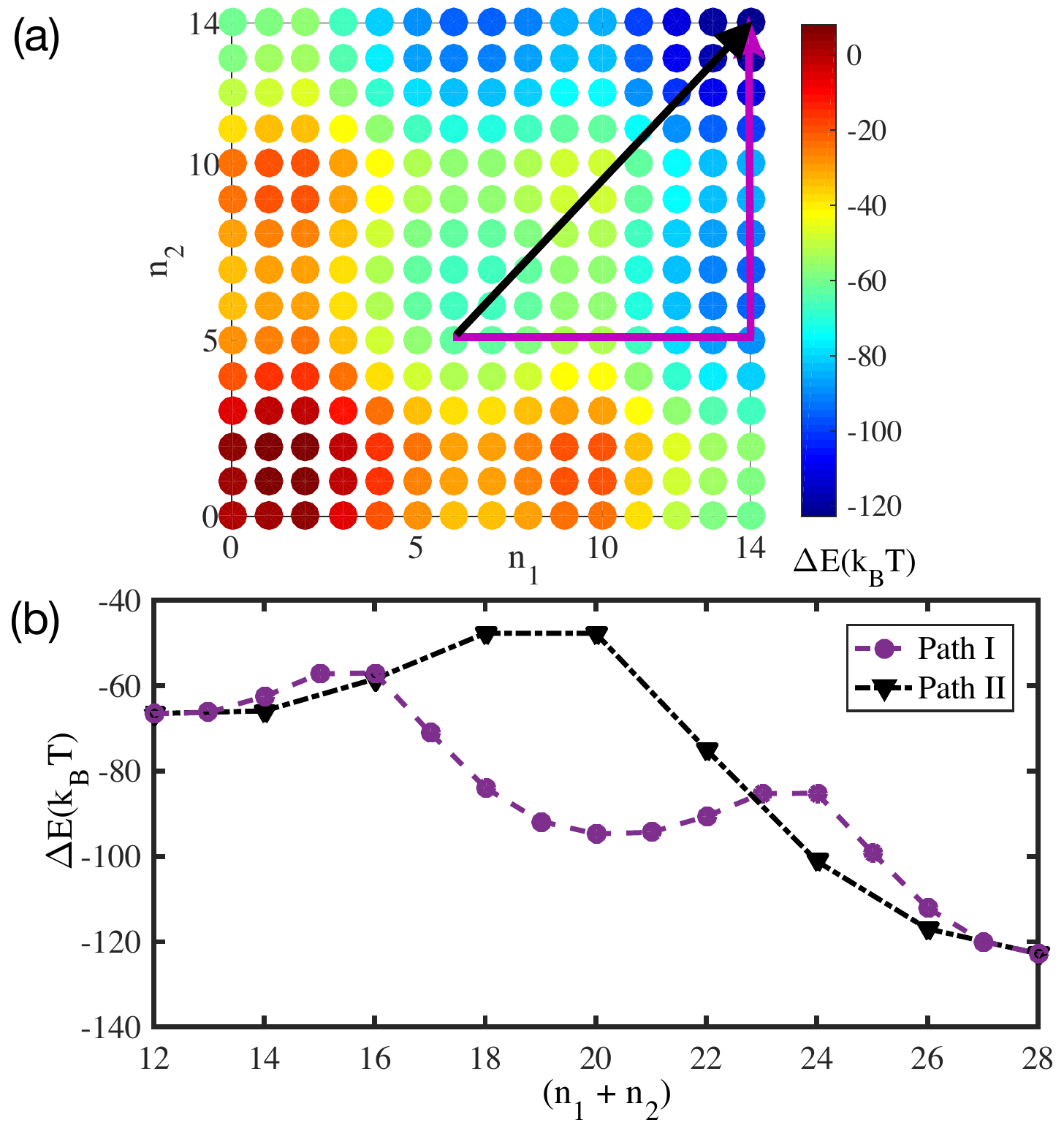}
 \caption{The free energy landscape of the dinucleosome, $\Delta E \equiv E_{min}(n_1,n_2) - E_{min}(n=0,n_2=0)$, versus the number of opened binding sites of the nucleosomes. The color bar in the right panel shows different colors corresponding to the energy in the unit of $k_BT$. Two possible paths for unwrapping are shown by the purple (path I) and black (path II) arrows. (b) $\Delta E$ versus the total number of opened binding sites, $n_1+n_2$, for the two paths shown in (a). These plots correspond to $F=8$ pN and $L_{linker} = 200$ bp.  }
\label{fig:En_landscape_F8} 	    
\end{figure}

In Figs. \ref{fig:En_landscape_F3}(a) and \ref{fig:En_landscape_F8}(a),  the energy landscape is depicted for two different forces. When the force is low, {\it i.e} $F= 1$ pN, the minimum of the energy is in the state of $n_1=n_2=0$, which means that the full wrapped nucleosomes remain stable. By increasing the force, the energy landscape changes, and new minima with different energy barriers appear. The studies on the mononucleosome reveal the nucleosome partial and full unwrapping happen in the forces of about $F\sim3$ pN and $F\sim8-9$ pN, respectively \cite{Laleh-2012,Mihardja-2006}. Figures \ref{fig:En_landscape_F3} and \ref{fig:En_landscape_F8} correspond to these forces. As indicated, increasing the external pulling force causes the displacement of the global minimum of the conformation energy, which indicates that the partial wrapped and full unwrapped nucleosomes are more preferable for the system. 

An important question is how the system gets to these points. To answer this question, we consider two paths that are characterized by colored arrows in the plots of Figs.  \ref{fig:En_landscape_F3}(a) and \ref{fig:En_landscape_F8}(a) for two forces of $F= 3$ pN and $F= 8$ pN, respectively. One of the typical paths runs along the $n_1$ and $n_2$ axes from the local minimum to the global minimum point (path $I$), which results in partial unwrapping of the second nucleosome after the first one. In the other path, the system reaches the global minimum through the diameter (path $II$), in which the number of opened points for each of the nucleosomes are equal. In Figs. \ref{fig:En_landscape_F3}(b) and \ref{fig:En_landscape_F8}(b) the energy of the system through these two paths has been shown for two different forces. In these figures, the behavior of $\Delta E \equiv E_{min}(n_1,n_2) - E_{min}(n_1=0,n_2=0)$ is depicted in terms of the total number of opened binding sites, $n_{tot}=n_1+n_2$, for two different forces. We note that in $F= 3$ pN and $F= 8$ pN, the global minima energy of the system correspond to the unwrapping of the first turn, and the full unwrapping, respectively. Furthermore, the larger energy barrier on the path $II$ reduces the unwrapping process along this path, according to the transition rates of Eqs. (\ref{eq:kuw1})-(\ref{eq:krw2}). Therefore, the probability of unwrapping across the path $I$ is higher. We emphasize that unwrapping of the nucleosomes is possible through the two mentioned paths, but the unwrapping is more probable through path I.  

\begin{figure}
 \centering
 \includegraphics[width=0.96\linewidth]{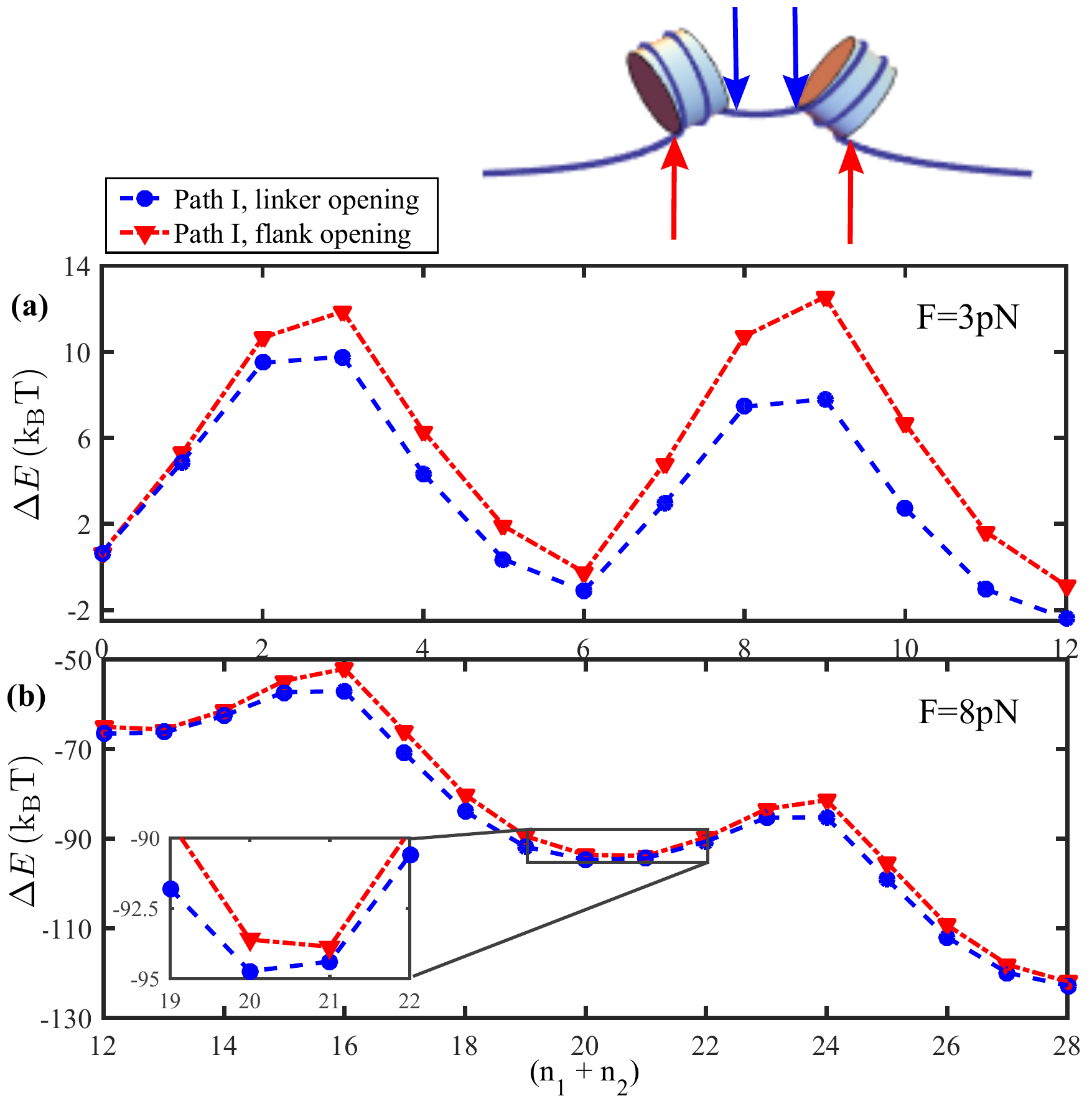}
 \caption{The free energy landscape of the dinucleosome, $\Delta E \equiv E_{min}(n_1,n_2) - E_{min}(n=0,n_2=0)$, versus the total number of opened binding sites, $n_1+n_2$. The triangle symbols (red dashed-dotted lines) correspond to the unwrapping of the nucleosomes from the linker DNA side, and solid circles (blue dashed lines) correspond to the case in which the unwrapping happens on the free DNA sides. (a) and (b) correspond to $F=3$ pN and $F=8$ pN, respectively. In both plots, $L_{linker} = 20$ bp. The schematic picture on the above panel, shows the places that the unwrapping occurs in two different situations. }
\label{fig:linker-flank-opening} 	    
\end{figure}

In Fig. \ref{fig:linker-flank-opening}, $\Delta E$ is depicted in terms of $n_{tot}=n_1+n_2$ for the asymmetric unwrapping of the nucleosomes for $L_{Linker}=20$ bp. As it is discussed above, for the short linker DNA, the energy landscape corresponding to the opening binding sites from the left and right of the nucleosomes is not symmetric. As an example, we consider an asymmetric unwrapping through path $I$. In Fig. \ref{fig:linker-flank-opening}, the triangle symbols (red dashed-dotted lines) correspond to the situation, where the nucleosomes are unwrapped from the linker DNA side, whereas solid circles (blue dashed lines) correspond to the case in which the unwrapping happens on the free DNA sides. As seen in this figure, there is a considerable difference in the energy of these two types of opening.

\begin{figure}
 \centering
 \includegraphics[width=0.9\linewidth]{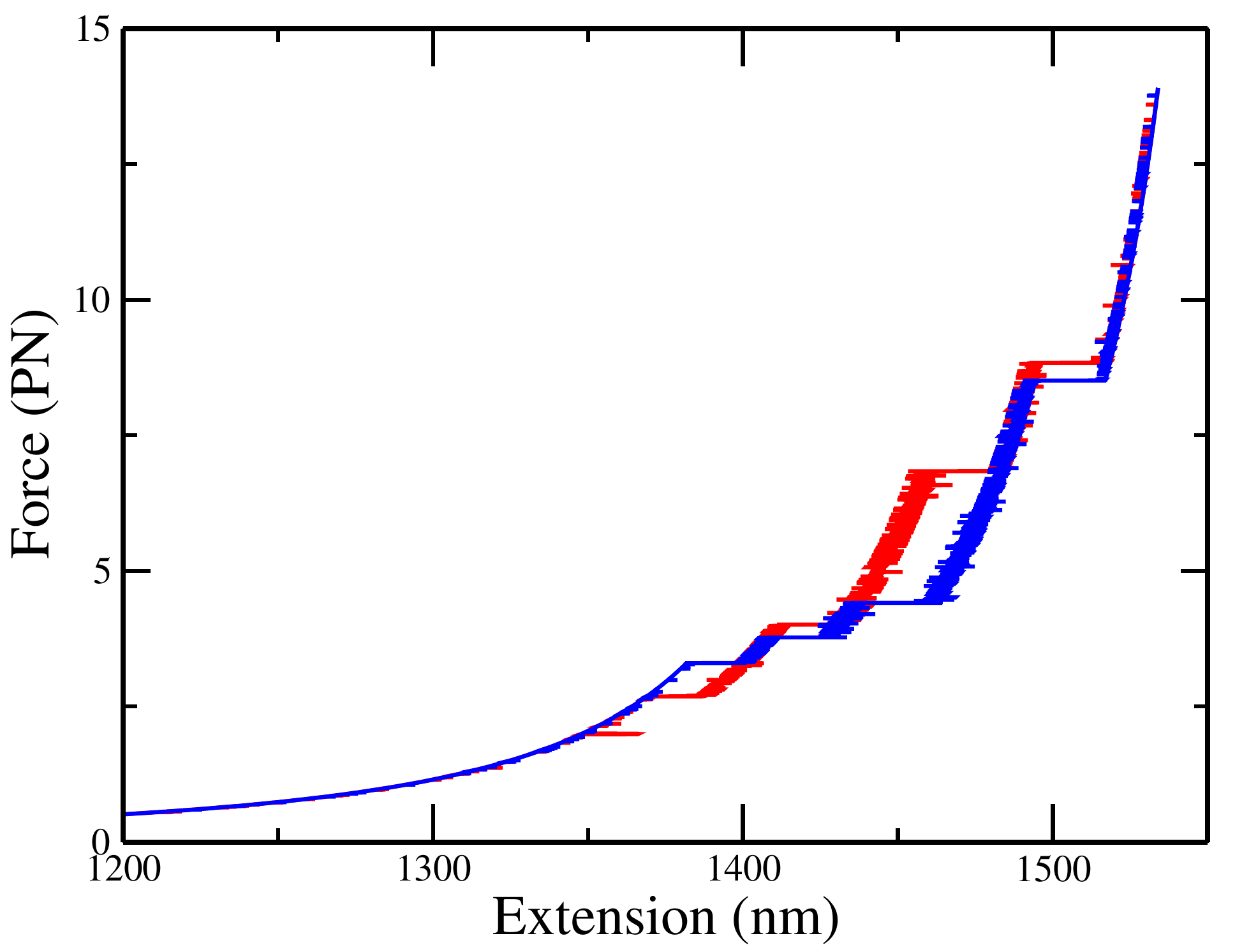}
 \caption{Two samples of force-extension curves for the unfolding of the dinucleosome with two long flanking DNA for a constant loading rate $k_{load}=2.4$ pN/s. }
 \label{fig:x_F} 	    
\end{figure}

\begin{figure}
 \centering
 \includegraphics[width=0.9\linewidth]{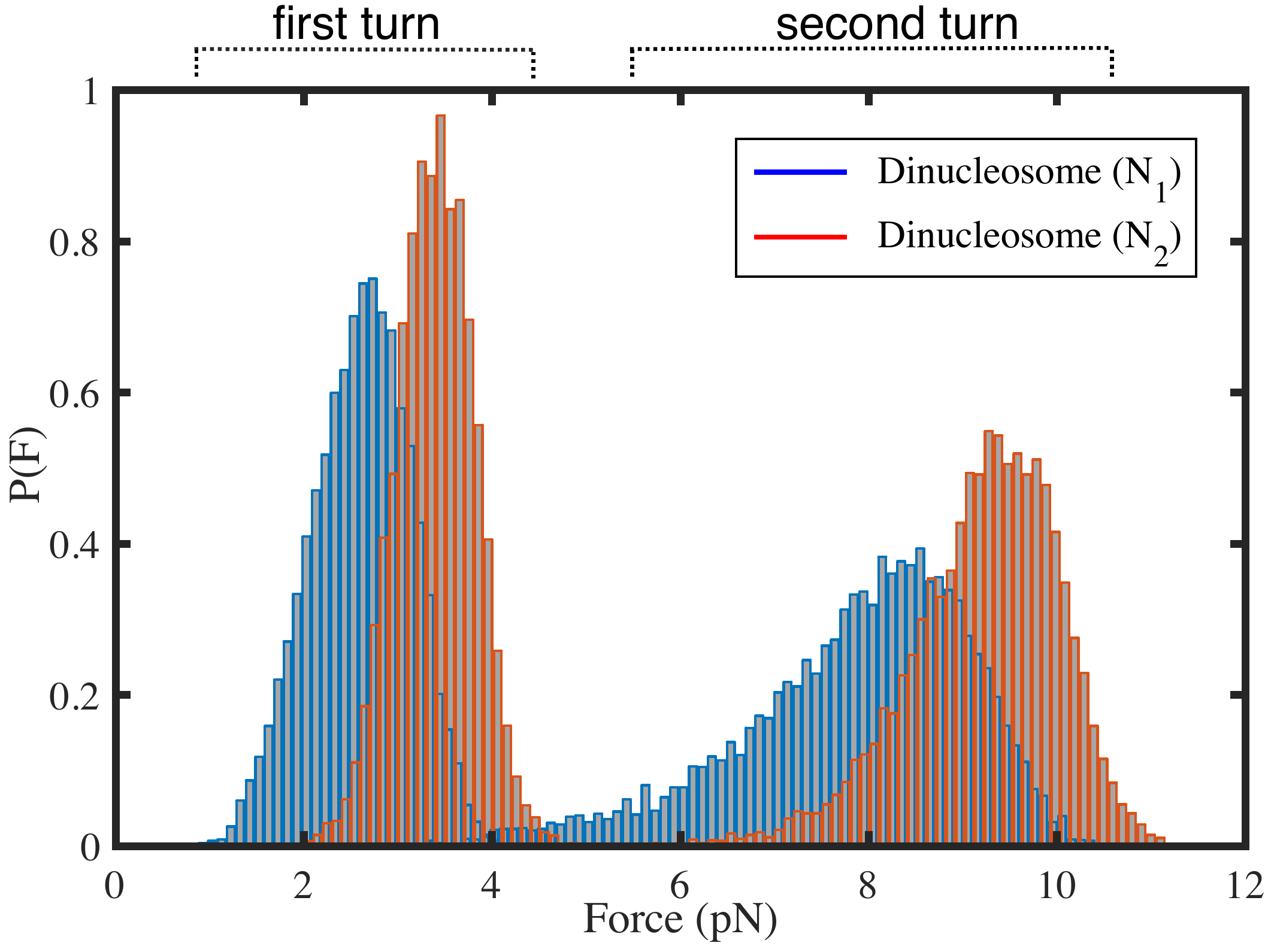}
 \caption{Histograms of force for the two abrupt transitions of the nucleosome. The black and red/blue lines display simulation for mononucleosome and dinucleosome respectively. For $F<5$ pN unwrapping of the outer turn happens and inner turn opens for $F>5$ pN. For simulation of dinucleosome, the initial length of linker DNA is $L_{linker}=200$ bp.}
\label{fig:compair_1Nuc_2Nuc_LMfixed} 
\end{figure}

To explore the dynamical features of the problem, we consider a possible experimental setup, in which a constant loading rate is applied on the two free ends of the dinucleosome. The force-extension curve obtained from the simulation has been shown in Fig. \ref{fig:x_F}. Four abrupt transitions in the length of DNA indicate the opening of the two turns of the nucleosomes. We note that for each experiment, the opening force can be different. When the force is low, both two nucleosomes prefer to be in the full wrapping state. When the stretching force increases, the first turn of the nucleosomal DNA of one of the nucleosomes is unwrapped.  After that, the next nucleosome is unpeeled partially. The larger the loading force gets, the more the probability of unwrapping of the nucleosomes becomes. To see this, Fig. \ref{fig:compair_1Nuc_2Nuc_LMfixed} shows the histogram of forces for the nucleosomes unwrapping corresponding to the constant loading rate $k_{load} = 2.44$ pN/s. The histogram corresponds to the constant loading rate case, therefore there is a small shift in the opening curves of the first turn and the second turn. The smaller the loading rate gets, the smaller the shift in the histograms of the opening becomes.


\section*{DISCUSSION AND CONCLUSION}  

We have used the force and torque balance conditions to find the equilibrium orientations of the dinucleosome in the presence of the stretching force, where the nucleosomes were allowed to unwrap. For a given external force, the system settled down to the state corresponding to the minimum energy. We have seen that the dinucleosome conformational energy is affected by the inter-nucleosomal spacing, which is due to the orientation of the nucleosomes. When the linker DNA length is short, the bending energy corresponding to the linker becomes more important, as seen in the conformations of Fig. \ref{fig:E_n1n2_F8_S}.   Therefore, by reducing the length of the linker DNA, the bending energy significantly increases and hence causes an asymmetric unwrapping of the dinucleosome, as discussed in the result section.

Asymmetric behavior has been also seen by Ngo {\it et al.} \cite{Ngo-2015}. It has been shown that the direction of the nucleosome unwrapping depends on the DNA flexibility: different flexibilities on both sides results in an asymmetric unwrapping of the DNA from the stiffer side,  while a stochastic opening occurs for equal flexibilities. This phenomenon and the observed asymmetric unwrapping in the dinucleosome, occur in a way that reduces the energy cost of the bent DNA. In experiments, different features of the unwinding such as the asymmetric and symmetric openings have been observed with different probabilities \cite{Ngo-2015,Chen-2014,Chen-2017,Halic-2018-NSMB}. The different patterns of the unwinding of the dinucleosome can be understood from the energy landscape, (see Fig. \ref{fig:linker-flank-opening}, and the transition rates of Eqs. (\ref{eq:kuw1})-(\ref{eq:krw2})). We note that a higher energy barrier creates less chance of occurrence. We have also applied a kinetic model to study the unwrapping dynamics of dinucleosome. The sequential unwrapping of the nucleosomes, as well as a delayed opening of the inner turns after the outer turns,  are observed which are in good agreement with the pulling experiments of the nucleosomal array  \cite{Wang-2002,vanNoort-NAR2015,Chien-2014}.

Comparison of dinucleosome and mononucleosome opening forces shows that the presence of the second nucleosome slightly reduces the amount of first nucleosome unwrapping forces (see Fig. \ref{fig:compair_1Nuc_2Nuc_LMfixed}). The reduction can be understood from the energy landscape of dinucleosome, as shown in figures \ref{fig:E_n1n2_F8_S}, \ref{fig:En_landscape_F3}, and \ref{fig:En_landscape_F8}. In the dinucleosome, increasing the degrees of freedom helps the nucleosome to find a path with lower energy barriers respect to the mono-nucleosome. The proposed path takes place along the path $I$ with fluctuations of one or two binding sites. This result is in good agreement with the observation of Fitz {\it et al.} on the dinucleosome template \cite{Fitz-2016}. 
They have seen a change in the transcription dynamics of the RNA polymerase II, as well as the variation of the opening forces in the force-extension experiment, in the dinucleosome compared to mono-nucleosome. They have reported that the reason for these changes is the presence of the second nucleosome, which confirms the effectiveness of this presence.

It is worth noting that the presence of the second nucleosome changes the dynamical features of a dinucleosme respect to a mono nucleosome, which can be understood by looking at the energy landscapes of each case. As it is discussed before, in the dinucleosome there are different possibilities for dynamics from the initial state to the final state. Each path corresponds to a different energy landscape, and the dynamical details such as the time needed for going from one state to the final state, depend on the path and its energy landscape. Therefore, the presence the the second nucleosome affects the energy landscape and may changes the dynamical aspects of the problem. This effect might be seen in the transcription rate of the nucleosomal DNA. It has been observed the transcription rate of RNA polymerase II through DNA becomes faster in the presence of two nucleosomes (dinucleosome case) than for mononucleosome \cite{Fitz-2016}.

Using cryo-EM, it has been observed that the stability of H2A-H2B dimers in the histone octamer depends on the presence of wrapped DNA in the nucleosome \cite{Halic-2018-NSMB}. Furthermore, the disassembly of the H2A-H2B dimers causes further DNA unwrapping \cite{Halic-2018-NSMB}. Possible extensions of this work may consider these effects and other conformational changes of the histone proteins in the partially wrapped nucleosomes and study the dinucleosome dynamics.

When the length of the linker DNA is long enough (say longer than ~ $30-40$ bp) , the presence of several nucleosomes does not affect on the energy landscape and one can consider the energy landscape of an array of $N$ nucleosomes as the energy landscape of $N$ single mono-nucleosomes. But for shorter lengths of the linker DNA (say shorter than ~$20-30$ bp), the presence of several nucleosomes affects the energy landscape. Therefore, for studying the dynamics of multi nucleosome systems, one should consider the effects of the length of the linker DNA. Generally, obtaining the energy landscape for nucleosome arrays with short linker DNA is not easy, and one should solve the force and torque balance equations, mentioned and discussed in this paper and ref. \cite{Hashem-Fatemeh1}.

\section*{ACKNOWLEDGMENTS}

We thank Maniya Maleki, Samaneh Rahbar, and Bahman Farnudi for very helpful comments on the manuscript.


\begin{thebibliography}{}

\bibitem{Cell}
B. Alberts {\it et al.}, {\it Molecular Biology of the Cell} (Garland, New York, 2007), 5th ed.

\bibitem{Widom-1997}
J. Widom, Curr. Biol. {\bf 7}, R653 (1997).

\bibitem{Khorasan-2004}
S. Khorasanizadeh, Cell {\bf 116}, 259 (2004).

\bibitem{Widom-2000}
J.D. Anderson, and J. Widom, J. Mol. Biol. {\bf 296} 979 (2000).

\bibitem{Flaus-2003}
A. Flaus, and T. Owen-Hughes, Biopolymers {\bf 68}, 563 (2003).

\bibitem{Widom-2005}
G. Li, M. Levitus, C. Bustamante, and J. Widom, Nat. Struct. Mol. Biol. {\bf 12}, 46 (2005). 

\bibitem{Bustamante-2009}
C. Hodges {\it et al.}, Science {\bf 325}, 626 (2009).

\bibitem{Logie-2009}
J.J.F.A. van Vugt {\it et al.}, PLoS ONE {\bf 4}, e6345 (2009). 

\bibitem{Bennink-2001}
M.L. Bennink {\it et al.}, Nat. Struct. Biol. {\bf 8}, 606 (2001). 

\bibitem{Wang-2002}
B.D. Brower-Toland {\it et al.}, Proc. Natl. Acad. Sci. U.S.A. {\bf 99}, 1960 (2002).

\bibitem{Bustamante-2000}
Y. Cui, and C. Bustamante, Proc. Natl. Acad. Sci. U.S.A. {\bf 97}, 127 (2000).

\bibitem{vanNoort-Nat2009}
M. Kruithof {\it et al.}, Nat. Struct. Mol. Biol. {\bf 16}, 534 (2009).

\bibitem{vanNoort-NAR2015}
H. Meng, K. Andresen, and J. van Noort, Nucl. Acids. Res. {\bf 43}, 3578 (2015). 

\bibitem{Mihardja-2006}
S. Mihardja, A. J. Spakowitz, Y. Zhang, and C. Bustamante, Proc. Natl. Acad. Sci. USA. {\bf 103}, 15871 (2006).

\bibitem{vanNoort-2009-2}
M. Kruithof, and J. van Noort, Biophys. J. {\bf 96}, 3708 (2009).

\bibitem{Mack-2012}
A.H. Mack {\it et al.}, J. Mol. Biol. {\bf 423}, 687 (2012).

\bibitem{Wang-2013}
M.Y. Sheinin {\it et al.}, Nat. Commun. {\bf 4}, 2579 (2013).

\bibitem{Wang-2009}
M.A. Hall {\it et al.}, Nat. Struct. Mol. Biol. {\bf 16}, 124 (2009).

\bibitem{Kulic-2004}
I. M. Kuli$\acute{c}$, and H. Schiessel, Phys. Rev. Lett. {\bf 92}, 228101 (2004).

\bibitem{Spakowitz-2011}
B. Sudhanshu {\it et al.}, Proc. Natl. Acad. Sci. U.S.A. {\bf 108}, 1885 (2011).

\bibitem{Laleh-2012}
L. Mollazadeh-Beidokhti, F. Mohammad-Rafiee, and H. Schiessel, Biophys. J. {\bf 102}, 2235 (2012).

\bibitem{Ngo-2015}
T.T.M. Ngo, Q. Zhang, R. Zhou, J.G. Yodh, and T. Ha, Cell {\bf 160}, 1135 (2015). 

\bibitem{Hashem-Fatemeh1}
H. Fatemi, F. Khodabandeh, and F. Mohammad-Rafiee, Phys. Rev. E {\bf 93}, 042409 (2016).

\bibitem{Nam-Arya-2014}
G.-M. Nam, and G. Arya, Nucl. Acids. Res. {\bf 42}, 9691 (2014).

\bibitem{Luger-1997}
K. Luger, A.W. Mader, R.K. Richmond, D.F. Sargent, and T.J. Richmond, Nature {\bf 389}, 251 (1997).

\bibitem{Richmond-2003}
T. J. Richmond and C. A. Davey, Nature {\bf 423}, 145 (2003).

\bibitem{Kolomeisky-Fisher-2000}
A.B. Kolomeisky, and M.E. Fisher, Physica A {\bf 279}, 1 (2000).

\bibitem{Hanggi-RMP-1990}
P. H\"anggi, P. Talkner, and M. Borkovec, Rev. Mod. Phys. {\bf 62}, 251 (1990).

\bibitem{Gillespie-1976}
D.T. Gillespie, J. Comput. Phys. {\bf 22}, 403 (1976).


\bibitem{Chen-2014}
Y. Chen, {\it et al.}, Nucl. Acids Res. {\bf 42}, 8767 (2014). 

\bibitem{Chen-2017}
Y. Chen, {\it et al.}, Proc. Natil. Acad. Sci. U.S.A. {\bf 114}, 334 (2017).

\bibitem{Halic-2018-NSMB}
S. Bilokapic1, M. Strauss, and M. Halic, Nat. Struct. Mol. Biol. {\bf 25}, 101 (2018).

\bibitem{Chien-2014}
F.T. Chien, and T. van der Heijden, Biophys. J. {\bf 107}, 373 (2014).

\bibitem{Fitz-2016}
V. Fitz {\it et al.}, Proc. Natil. Acad. Sci. U.S.A. {\bf 113}, 12733 (2016).
\end{thebibliography}
\end{document}